\documentclass[aps,preprint,amsmath,amssymb,longbibliography,superscriptaddress]{revtex4-1}

\usepackage{graphicx}

\usepackage{hyperref}
\usepackage{bm}
\usepackage{color}
\usepackage{mathtools}
\usepackage{hyperref}
\hypersetup{colorlinks=true}
\usepackage{CJKutf8}

\begin{document}

%\preprint{}


\title{YUNIC: A Multi-Dimensional Particle-In-Cell Code for Laser-Plasma Interaction}

\affiliation{Beijing National Laboratory for Condensed Matter Physics, Institute of Physics, Chinese Academy of Sciences, Beijing 100190, China}
\affiliation{Department of Physics and Beijing Key Laboratory of Opto-electronic Functional Materials and Micro–nano Devices, Renmin University of China, Beijing 100872, China}
\author{Huai-Hang Song}
%\email{hhsong@iphy.ac.cn}
\affiliation{Beijing National Laboratory for Condensed Matter Physics, Institute of Physics, Chinese Academy of Sciences, Beijing 100190, China}
\author{Wei-Min Wang}
\affiliation{Department of Physics and Beijing Key Laboratory of Opto-electronic Functional Materials and Micro–nano Devices, Renmin University of China, Beijing 100872, China}
\author{Yu-Tong Li}
\affiliation{Beijing National Laboratory for Condensed Matter Physics,
Institute of Physics, Chinese Academy of Sciences, Beijing 100190, China}

\date{\today}

\begin{abstract}

For simulating laser-plasma interactions, we developed a parallel, multi-dimensional, fully relativistically particle-in-cell (PIC) code, named {\scshape yunic}. The core algorithm is introduced, including field solver, particle pusher, field interpolation, and current interpolation. In addition to the classical electromagnetic interaction in plasmas, nonlinear Compton scattering and nonlinear Breit-Wheeler pair production are also implemented based on Monte-Carlo methods to study quantum electrodynamics (QED) processes. We benchmark {\scshape yunic} against theories and other PIC codes through several typical cases. Y{\scshape unic} can be applied in varieties of physical scenes, from relativistic laser-plasma interactions to astrophysical plasmas and strong-field QED physics.

\end{abstract}

%\keywords{}

\maketitle
\tableofcontents
%\section{}
%% Put \label in argument of \section for cross-referencing
%%\section{\label{}}
%\subsection{}
%\subsubsection{}

\section{Introduction}

The particle-in-cell (PIC) method \cite{birdsall2004plasma}, belonging to a kinetic description, can accurately simulate the collective plasma behavior, from linear to relativistically nonlinear processes. Compared to magnetohydrodynamics simulation, PIC simulation utilizing the quasi-particle concept can resolve plasma dynamics on smaller spatial and temporal scales. Meanwhile, it requires much less computational expense than that by directly solving Vlasov-Boltzmann equations. Since it was initially developed in 1970s \cite{Dawson1983rmp}, PIC method has become one of the most powerful and indispensable tools in various plasma areas, particularly in laser-plasma interactions \cite{gibbon2005laserplasma,macchi2013laserplasma}. In this manuscript, we introduce a recently developed PIC code {\scshape yunic}, and demonstrate a few typical benchmarks to validate this code. Besides the classical plasma dynamics, {\scshape yunic} is also capable of simulating extremely laser-plasma interactions in the quantum electrodynamics (QED) regime, including spin and polarization effects in the processes of nonlinear Compton scattering and nonlinear Breit-Wheeler pair production \cite{Baier1998qed}.

\section{Standard particle-in-cell algorithm}
\label{algorithm}

The core idea of PIC method is to solve Maxwell’s equations on the discrete spatial grid [Sec.~\ref{field_solver}], while pushing quasi-particles in the free space [Sec.~\ref{particle_pusher}]. The currents generated by moving charged particles should be interpolated to the spatial grid as sources to solve field equations [Sec.~\ref{current_interpolation}], and fields should also be interpolated back to an arbitrary particle position to push them [Sec.~\ref{field_interpolation}]. Hence, the discrete fields and non-discrete particles are self-consistently connected. The main equations for solving collisionless plasma problems are as follows (in Gaussian units):
\begin{eqnarray}
&&\nabla\cdot{\bf E}=4\pi\rho, \label{eqq1}
\\
&&\nabla\cdot{\bf B}=0, \label{eqq2}
\\
&&\nabla\times{\bf E}=-\frac{1}{c}\frac{\partial\bf B}{\partial t}, \label{eqq3}
\\
&&\nabla\times{\bf B}=\frac{1}{c}\left(4\pi{\bf J}+\frac{\partial\bf E}{\partial t}\right), \label{eqq4}\\
&&\frac{\partial\rho}{\partial t}+\nabla\cdot{\bf J}=0, \label{eqq5}\\
&&\frac{d{\bf p}}{dt}=q({\bf E}_p+\frac{\bf p}{\gamma mc}\times{\bf B}_p), \label{eqq6}\\
&&\frac{d{\bf r}_p}{dt}=\frac{\bf p}{\gamma m},\label{eqq7}
\end{eqnarray}
where Eqs.~\eqref{eqq1}-\eqref{eqq4} are Maxwell's equations, Eq.~\eqref{eqq5} is the charge conservation equation, Eqs.~\eqref{eqq6} and \eqref{eqq7} are Newton-Lorentz equations.

\subsection{Normalization}\label{field_solver}
In the PIC code, after defining a reference frequency $\omega_r$, it is convenient to normalize time $t$, length $L$, electric field $\bf E$, magnetic field $\bf B$, particle velocity $\bf v$, momentum $\bf p$, number density $n$, current density $\bf J$, charge $q$ and mass $m$ to following quantities, respectively:
\begin{eqnarray}
t_r&=&\frac{1}{\omega_r}, \quad L_r=\frac{c}{\omega_r},\quad E_r=\frac{m_e\,c\,\omega_r}{e}, \quad B_r=\frac{m_e\,c\,\omega_r}{e}, \quad
v_r=c,\label{eqq8}\\ p_r&=&m_e\,c, \quad n_r=\frac{m_e\,\omega_r^2}{4\pi e^2}, \quad J_r=c\,e\,n_r, \quad q_r=e, \quad m_r=m_e,\nonumber
\end{eqnarray}
where $m_e$ is the electron rest mass, and $e$ is the electron charge. In laser-plasma interactions, $\omega_r$ is usually chosen as the laser frequency $\omega_L$, and it could also be chosen as the plasma oscillating frequency $\omega_{p}$ in other interactions. For convenience, we use normalized quantities below based on Eq.~\eqref{eqq8}.

\subsection{Field solver}\label{field_solver}

\begin{figure}[t]
\centering
\includegraphics[width=5.5in]{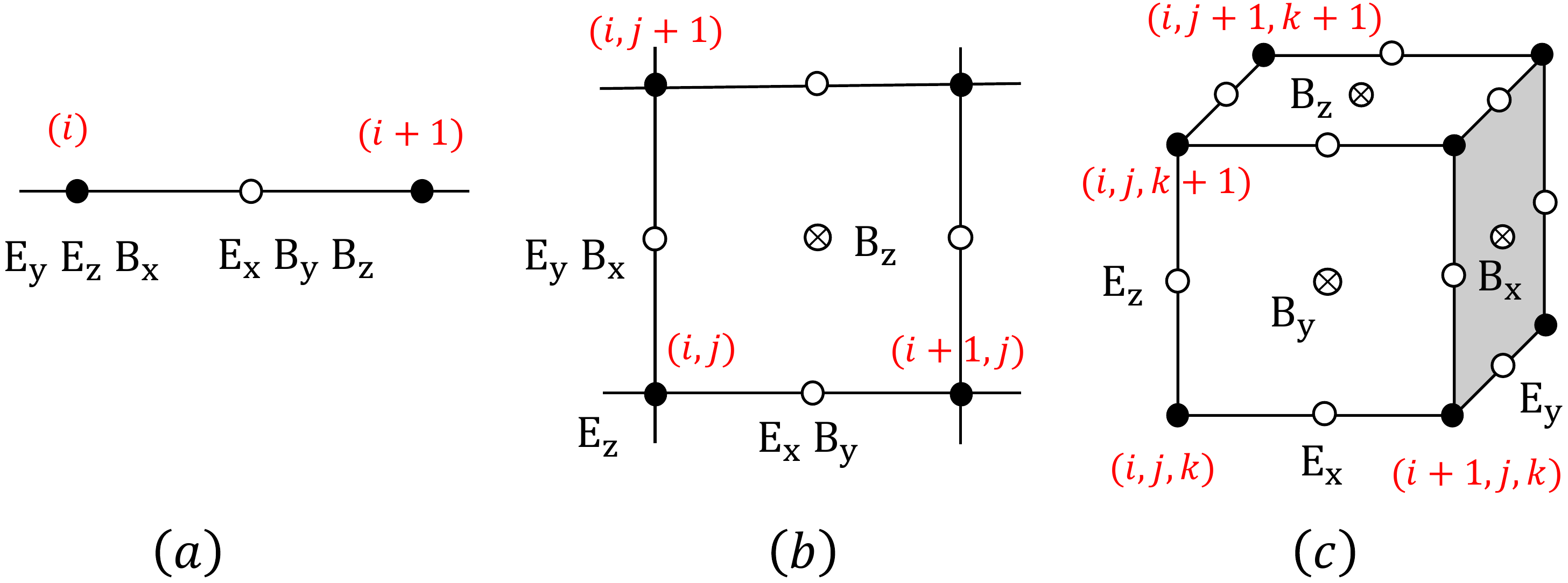}
\caption{\label{fig1} The staggered Yee grid in (a) one dimensional (1D), (b) two dimensional (2D), and (c) three dimensional (3D) spaces for solving Maxwell's equations, respectively.}
\end{figure}

The electromagnetic fields ${\bf E}$ and ${\bf B}$ are self-consistently evolved by solving Maxwell's equations. One only needs to solve two curl equations [Eqs.~\eqref{eqq3} and \eqref{eqq4}] and a charge conservation equation [Eq.~\eqref{eqq5}], because the other two divergence equations [Eqs.~\eqref{eqq1} and \eqref{eqq2}] are automatically satisfied with time if they hold initially \cite{EASTWOOD1991252,Villasenor1992cpc}. The details of how to realize the charge conservation are presented in Sec.~\ref{current_interpolation}.

The finite-difference time-domain (FDTD) method \cite{Taflove2005book} is adopted for numerically solving Maxwell's equations. Equations.~\eqref{eqq3} and \eqref{eqq4} can be written as the following discrete forms with the 2nd-order accuracy:
\begin{eqnarray}
{\bf E}^{n+1/2}-{\bf E}^{n-1/2}&=&\Delta t(\nabla\times{\bf B}^n-{\bf J}^n), \label{eqq9}
\\
{\bf B}^{n+1}-{\bf B}^{n}&=&-\Delta t\nabla\times{\bf E}^{n+1/2}. \label{eqq10}
\end{eqnarray}
Here, the leapfrog scheme is adopted in time and the famous staggered Yee grid \cite{Yee1966ieee} is employed in space, as illustrated in Fig.~\ref{fig1}.

\subsection{Particle pusher}\label{particle_pusher}
We proceed to solve the motion of charged particles in the electromagnetic field by discretizing Eqs.~\eqref{eqq6} and ~\eqref{eqq7} as following:
\begin{eqnarray}
\frac{{\bf u}^{n+1}-{\bf u}^{n}}{\Delta t}&=&q_m({\bf E}_p^{n+1/2}+\frac{1}{{\bf\gamma}^{n+1/2}}{\bf u}^{n+1/2}\times{\bf B}_p^{n+1/2}) \label{eqq11},
\\
\frac{{\bf r}_p^{n+3/2}-{\bf r}_p^{n+1/2}}{\Delta t}&=& \frac{{\bf u}^{n+1}}{\gamma^{n+1}}\label{eqq12},
\end{eqnarray}
where ${\bf u}={\bf p}/m$ and $q_m=q/m$.

\begin{figure}[t]
\centering
\includegraphics[width=3.in]{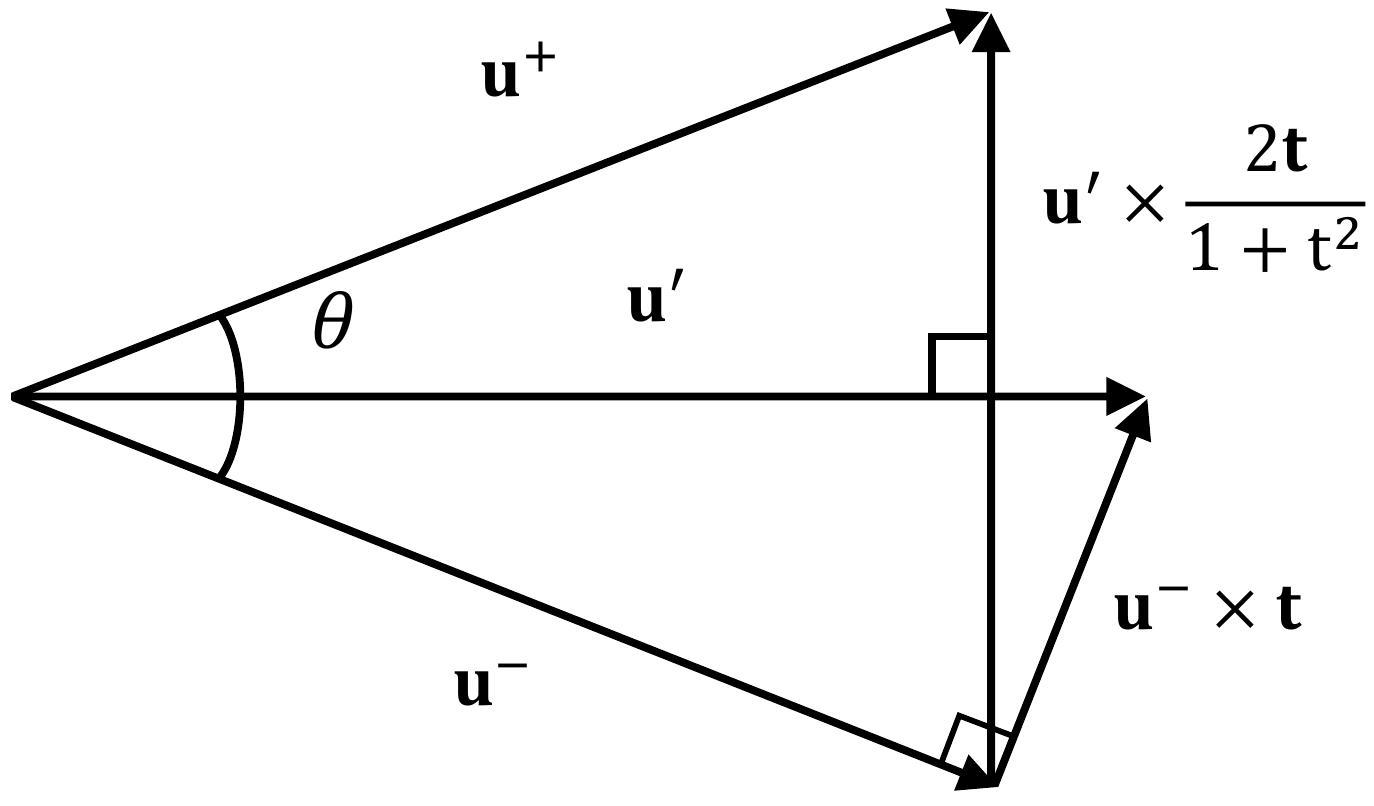}
\caption{\label{fig2} Diagram of Boris algorithm.}
\end{figure}

Boris algorithm \cite{Boris1970conf} is employed to push charged particles due to its advantage of long term accuracy \cite{Qin2013pop}, which splits the electric and magnetic forces by defining ${\bf u}^{+}$ and ${\bf u}^{-}$,
\begin{eqnarray}
{\bf u}^{n}&=&{\bf u}^{-}-\frac{q_m\Delta t}{2}{\bf E}_p^{n+1/2} \label{eqq13},\\
{\bf u}^{n+1}&=&{\bf u}^{+}+\frac{q_m\Delta t}{2}{\bf E}_p^{n+1/2} \label{eqq14}.
\end{eqnarray}
Substituting Eqs.~\eqref{eqq13} and \eqref{eqq14} into Eq.~\eqref{eqq11}, one can obtain a rotation equation of $\bf u^+$ and $\bf u^-$ about the magnetic field ${\bf B}_p$, i.e., $2\gamma^{n+1/2}({\bf u}^{+}-{\bf u}^{-})=q_m{\Delta t}({\bf u}^{+}+{\bf u}^{-})\times {\bf B}_p^{n+1/2}$, and then solve it by the following implementation:
\begin{eqnarray}
{\bf u}'&=&{\bf u}^{-}+{\bf u}^{-}\times{\bf t},\\
{\bf u}^{+}&=&{\bf u}^{-}+{\bf u}'\times\frac{2 {\bf t}}{1+t^2}
\end{eqnarray}
where ${\bf t}=q_m\Delta t{\bf B}_p^{n+1/2}/{(2\gamma^{n+1/2})}$. The diagram of Boris algorithm is illustrated in Fig.~\ref{fig2}.

If contributions of electric field and magnetic field to the Lorentz force nearly cancel out, e.g., in the rest frame of an ultrarelativistic beam \cite{Vay2007prl}, Boris algorithm might lead to large errors, and Vay algorithm \cite{Vay2008pop} is more suitable.

\subsection{Field interpolation}\label{field_interpolation}

\begin{figure}[t]
\centering
\includegraphics[width=6.1in]{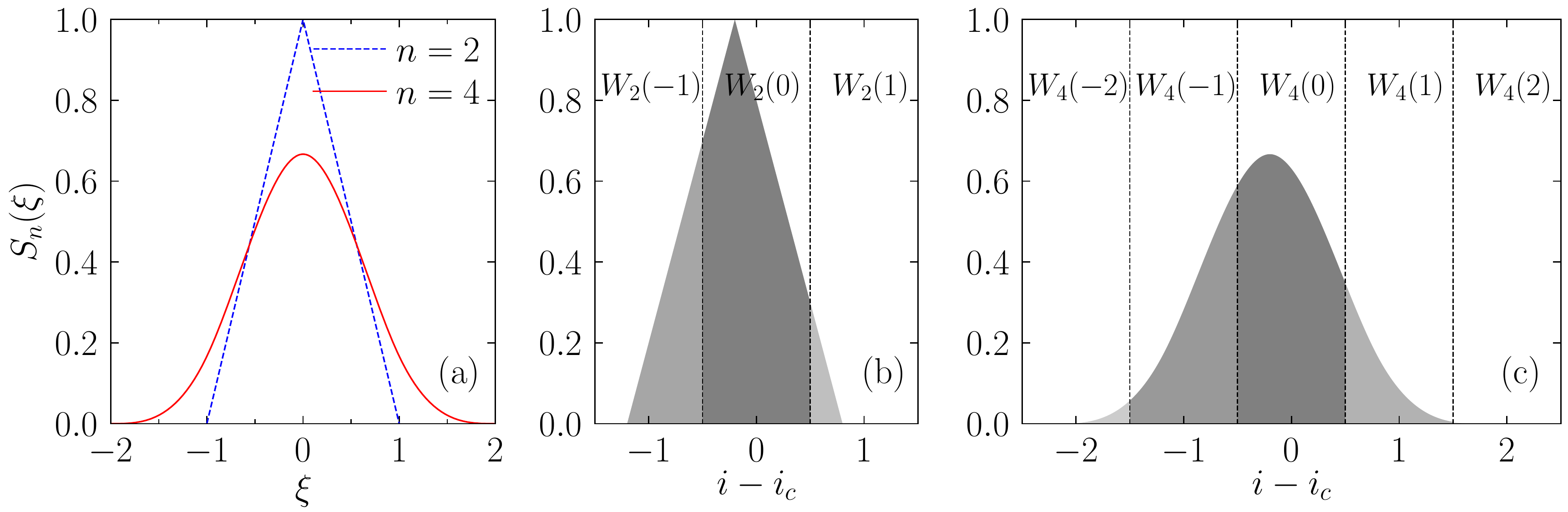}
\caption{\label{fig3} (a) Shape functions $S_n(\xi)$ of 2nd-order ($n=2$) and 4th-order ($n=4$) used in PIC for field and current interpolations. (b) (c) The area of the shadow $W_n(i-i_c)$ represents the weight of the grid field acting on the particle for $n=2$ and $n=4$, respectively. Normally, weight function $W_2(i-i_c)$ crosses three grid cells, and $W_4(i-i_c)$ crosses five grid cells.}
\end{figure}

In Sec.~\ref{particle_pusher}, we have discussed how to push a charged particle provided we have known the electric field ${\bf E}_p$ and magnetic field ${\bf B}_p$ at the particle position ${\bf r}_p(x_p,y_p,z_p)$. In this section, we discuss how to obtain ${\bf E}_p$ and ${\bf B}_p$ through the interpolation. Actually, it depends on the interpolation shape function $S_n(\xi)$ we choose, where $n$ is the interpolation order. The function $S_n(\xi)$ defines the shape and smoothness of quasi-particles and also determines the simulation accuracy \cite{birdsall2004plasma}. Taking 1D as an example, its 2nd-order and 4th-order forms shown in Fig.~\ref{fig3}(a) are given by \cite{Abe1986jcp}
\begin{eqnarray}
S_2(\xi)&=&\begin{dcases}
1-|\xi| &\text{if}\,\, |\xi|\le1,\\
0 &\text{otherwise},
\end{dcases}\\\nonumber
\\
S_4(\xi)&=&\begin{dcases}
\frac{2}{3}-|\xi|^2+\frac{1}{2}|\xi|^3 &\text{if}\,\, |\xi|\le1,\\
\frac{4}{3}(1-\frac{1}{2}|\xi|)^3 &\text{if}\,\, 1<|\xi|\le2,\\
0 &\text{otherwise}.
\end{dcases}
\end{eqnarray}
For a particle located at $x_p$, its acting field contributed by grid point $i$ can be expressed as $F(i)*W_n(i-i_c)$, where its weight function is defined by $W_n(i-i_c)=\int_{i-1/2}^{i+1/2}S_n(\xi-x_p/\Delta x)d\xi$ and $i_c$ is the grid point nearest to the particle, as shown in Figs.~\ref{fig3}(b) and \ref{fig3}(c). The expression of $W_n(i-i_c)$ after the integration can be found from APPENDIX A of \cite{Abe1986jcp}.
In 3D, the electromagnetic field acting on the particle can be calculated through following interpolations under the condition that the fields are constant over each cell:
\begin{eqnarray}
{\bf E}({\bf B})_p(x_p,y_p,z_p)=\sum_{i,j,k}W_n(i-i_c)*W_n(j-j_c)*W_n(k-k_c)*{\bf E}({\bf B})(i,j,k).
\end{eqnarray}
Notice that different field components generally have different weights since the grid is staggered [see Fig.~\ref{fig1}].
%, i.e., $i_c={\rm round}(x_p/\Delta x)$ or ${\rm round}(x_p/\Delta x-0.5)$ for staggered grids

\subsection{Current interpolation}\label{current_interpolation}
As we mentioned in Sec.~\ref{field_solver}, one needs to ensure the charge conservation in order to avoid solving Poisson's equation [Eq.~\eqref{eqq1}] \cite{Villasenor1992cpc}, since the local computation of the former is much simpler and computationally cheaper than the global computation of the latter. In {\scshape yunic}, Esirkepov algorithm  \cite{ESIRKEPOV2001144} is adopted to ensure the charge conservation in the current calculation. By assuming the particle trajectory over one time step is a straight line, the current flux in 3D is decomposed into twelve segments along $x$, $y$, and $z$ axes, respectively, i.e.,
\begin{eqnarray}
{\rm J}_x^{n+1}(i+1,j,k)-{\rm J}_x^{n+1}(i,j,k)&=&-q\frac{\Delta x}{\Delta t}W_x(i,j,k),
\\
{\rm J}_y^{n+1}(i,j+1,k)-{\rm J}_y^{n+1}(i,j,k)&=&-q\frac{\Delta y}{\Delta t}W_y(i,j,k),
\\
{\rm J}_z^{n+1}(i,j,k+1)-{\rm J}_z^{n+1}(i,j,k)&=&-q\frac{\Delta z}{\Delta t}W_z(i,j,k),
\end{eqnarray}
where $W_x(i,j,k)$, $W_y(i,j,k)$ and $W_z(i,j,k)$ can be found from Eq. (31) in \cite{ESIRKEPOV2001144}.

This method is easily extended to an arbitrary high-order shape function $S_n(\xi)$. Note that $S_n(\xi)$ employed in the current interpolation should be the same as that in the field interpolation [Sec.~\ref{field_interpolation}] to eliminate
the self-force. 

\begin{figure}[t]
\centering
\includegraphics[width=5.7in]{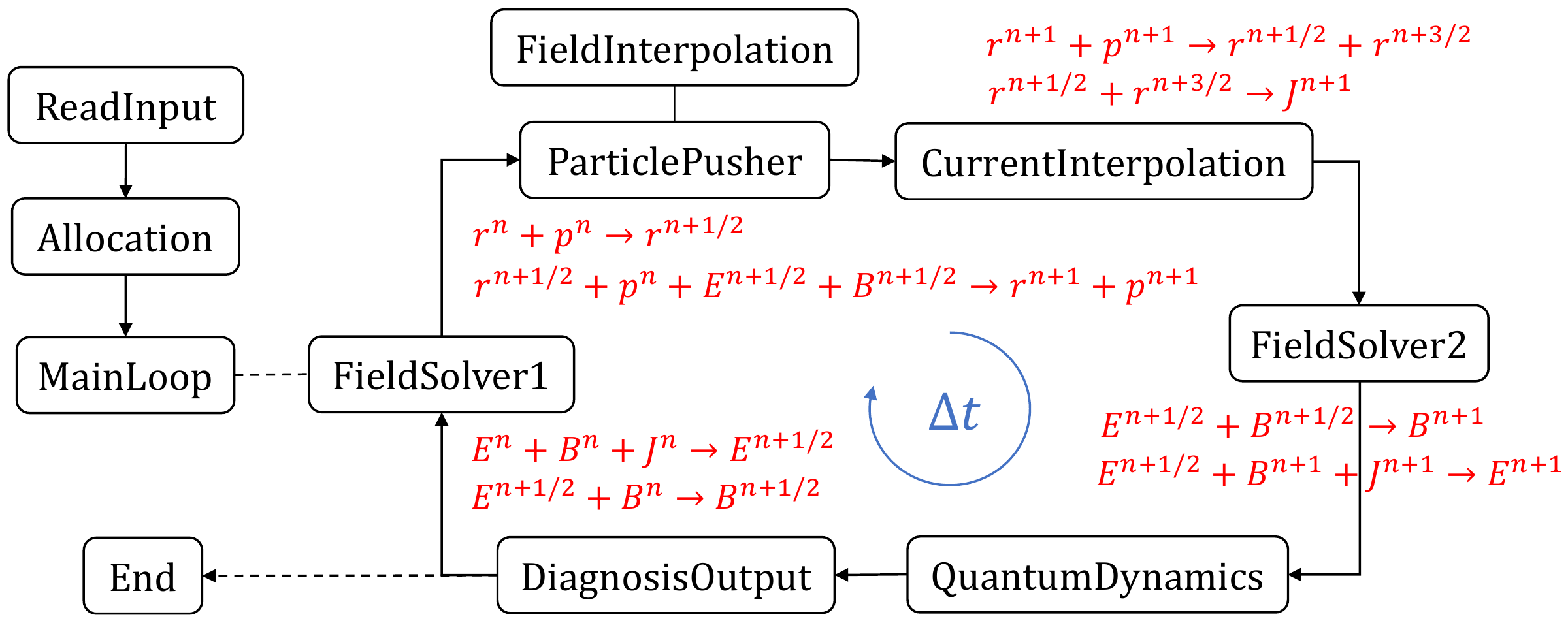}
\caption{\label{fig4} Flow chart of the core algorithm in {\scshape yunic}.}
\end{figure}

\subsection{Algorithm structure}\label{algorithm_structure}

{\scshape Yunic} is written in C++ language and massively parallelized by MPI. The simulation results are output in parallel by MPI-IO and then analyzed/visualized by Python/Matplotlib. Its three versions aimed at different spatial dimensions (1D, 2D, and 3D) are constructed separately for efficiency. {\scshape Yunic} employs a modified algorithm that originally adopted in {\scshape psc} \cite{Ruhl_Docu} by Hartmut Ruhl, and latter also adopted in {\scshape epoch} \cite{Arber2015ppcf}, which is slightly different from the common algorithm as described in Secs.~\ref{field_solver}-\ref{current_interpolation}. The modified algorithm updates the fields and particle positions at both full-time steps and half-time steps, as sketched in Fig.~\ref{fig4}. Hence, the drawback of the leapfrog method is overcome and one can obtain the information of fields and particles at the same time, which is important in some cases, e.g., for simulating QED processes.

\section{Benchmarks in several typical cases}

Here, we benchmark our PIC code {\scshape yunic} against the open-source PIC code {\scshape smilei} \cite{Derouillat2018cpc}, including 1D, 2D and 3D versions. At these presented cases, the simulation results of two PIC codes are in good agreement.

\begin{figure}[t]
\centering
\includegraphics[width=5.3in]{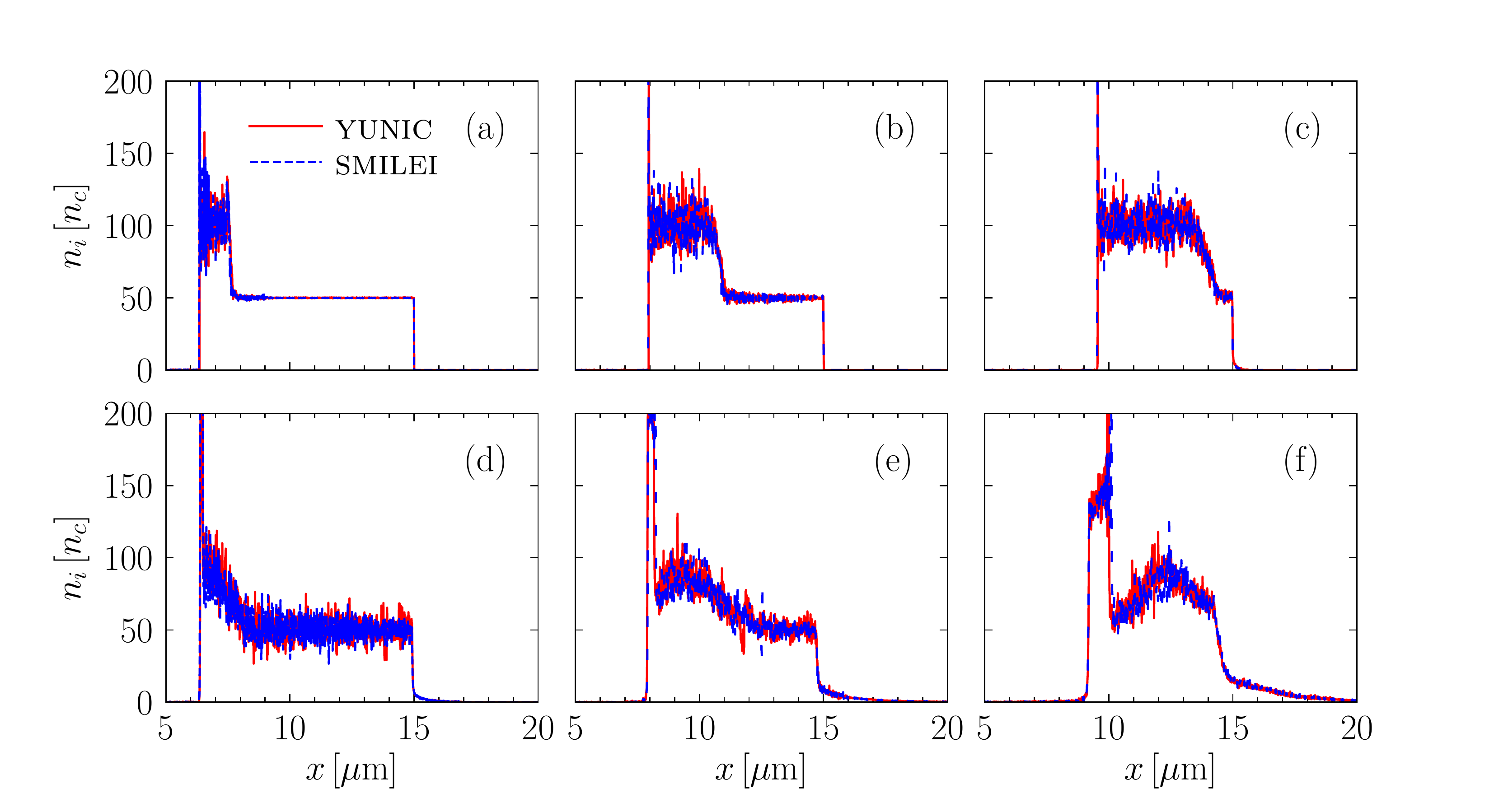}
\caption{\label{fig5} Spatial distributions of ion density $n_i$ at different times, (a)(d) $t=0.17$ ps, (b)(e) $t=0.33$ ps, and (c)(f) $t=0.5$ ps. (a)-(c) and (d)-(f) correspond to circularly and linearly polarized lasers, respectively.}
\end{figure}

(a) \emph{1D PIC simulation: hole boring}. A linearly (circularly) polarized laser with a wavelength of $\lambda_L=1~\mu\rm m$ incidents from the left boundary at $t=0$ ps. The laser has a 5-laser-period rise time before a constant normalized intensity of $a_0=eE_L/m_ec\omega_L=10{\sqrt 2}~(10)$. An uniform overdense plasma with an electron density of $n_p=50n_c$ is located at $5~\mu{\rm m}<x<15~\mu{\rm m}$, where $n_c=m_e\omega_L^2/4\pi e^2$. The computational domain has a size of $20\lambda_L$ with 5120 cells in the $x$ direction. Each cell contains 100
electrons and 100 protons. Absorbing boundaries are used for both particles and fields. The 4th-order interpolation is employed. The comparison of ion charge density $n_i$ at different times are shown in Fig.~\ref{fig5}.

\begin{figure}[t]
\centering
\includegraphics[width=3.3in]{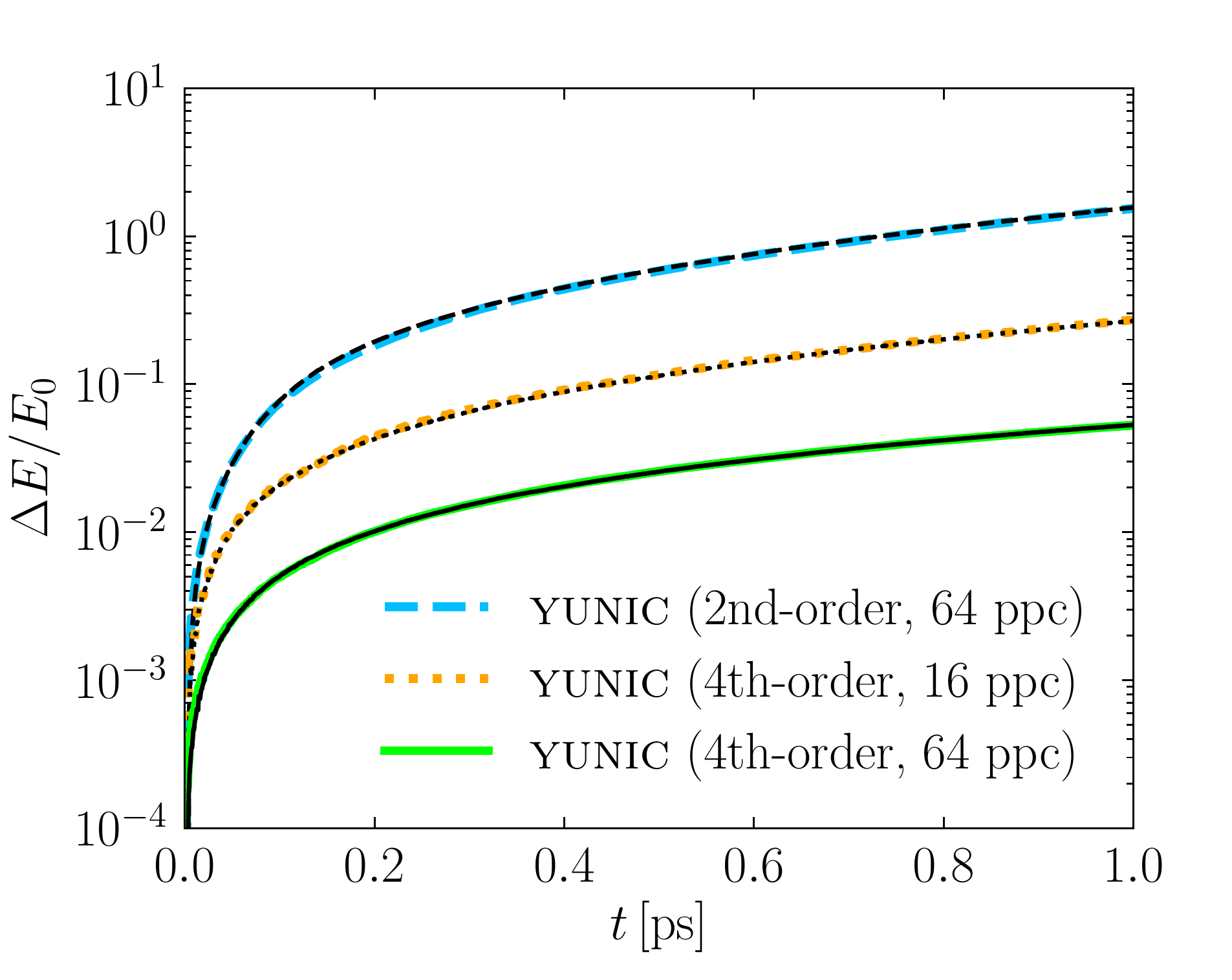}
\caption{\label{fig6} Time evolution of the relative increase of total energy $\Delta E/E_0$ with different interpolation orders and particles per cell (ppc). The colored thick lines are the simulation results by {\scshape yunic}, and black thin lines correspond to those by {\scshape smilei}.}
\end{figure}

\begin{figure}[t]
\centering
\includegraphics[width=6.4in]{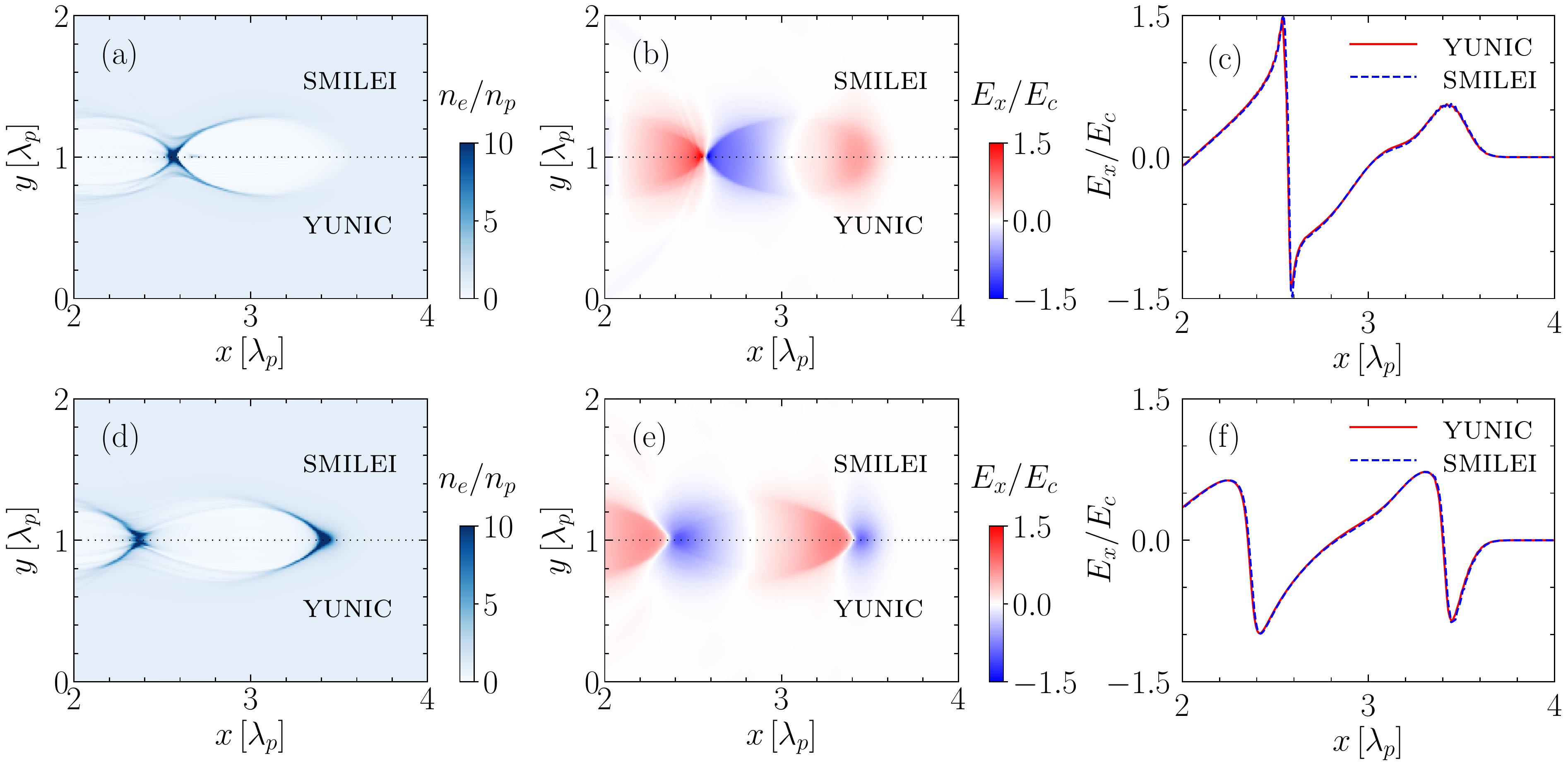}
\caption{\label{fig7} (a)(d) Electron density $n_e$ of the background plasma. (b)(e) Longitudinal electric field $E_x$. (c)(f) On-axis distribution of $E_x$. (a)-(c) and (d)-(f) are driven by electron and positron beams, respectively.}
\end{figure}

(b) \emph{2D PIC simulation: self-heating}. A plasma with an initial temperature of 1 keV has an uniform electron density of $n_p=100n_c$. Each plasma wavelength ($\lambda_p=2\pi/k_p$) contains 320 grids. Periodic boundaries are applied for both particles and fields. No current smoothing or other additional algorithms are used to control self-heating. The comparison of relative energy increasing $\Delta E/E_0$ due to self-heating are shown in Fig.~\ref{fig6}.

(c) \emph{3D PIC simulation: wakefield driven by electron or positron beam}. A 1-GeV drive electron (positron) beam has a bi-Gaussian density profile with a transverse size $k_p\sigma_r=0.5$, bunch length $k_p\sigma_x=0.5$, and peak density $n_b/n_p=4$. The computational domain has a size of $4\lambda_p\times 2\lambda_p\times 2\lambda_p$ in $x\times y\times z$ directions, sampled by $512\times 192\times 192$ cells. Each cell contains 8 electrons and 8 protons for the uniform background plasma and 4 electrons or positrons for the drive beam. For the field initialization of an ultrarelativistic charged beam, {\scshape yunic} first solves the Poisson's equation in the beam's rest frame, and then applies Lorentz transformation to obtain its self-generated fields in the laboratory frame \cite{Massimo2016}. The comparison of background plasma density $n_e$ and excited longitudinal electric field $E_x$ are shown in Fig.~\ref{fig7}.

\section{QED modules}

The available petawatt and next-generation 10-petawatt and 100-petawatt laser systems can provide an extreme field density of $I_0=10^{22-25}\rm W/cm^2$. To explore interactions of ultraintense lasers with plasmas, {\scshape yunic} have implemented QED modules with Monte-Carlo methods \cite{Elkina2011prab,Ridgers2014jcp,Gonoskov2015pre} to calculate the photon emission via nonlinear Compton scattering and the electron-positron pair production via nonlinear Breit-Wheeler process.

%\subsection{Photon emission}
%\label{photon_emission}

The spin- and polarized-averaged differential  probability of the photon emission is written as \cite{Baier1998qed}
\begin{eqnarray}
\frac{d^2W_{\rm rad}}{dudt}=\frac{\alpha m^2c^4}{\sqrt{3}\pi \hbar \varepsilon_e}\left[\frac{u^2-2u+2}{1-u}K_{2/3}(y)-\int_{y}^{\infty}K_{1/3}(x)dx \right]\label{eqq23},
\end{eqnarray}
where $K_{\nu}(y)$ is the second-kind $\nu$-order modified Bessel function, $y=2u /[3(1-u)\chi_e]$, $u=\varepsilon_\gamma / \varepsilon_e$, $\varepsilon_e$ is the electron energy before the photon emission, $\varepsilon_\gamma$ is the emitted photon energy, and $\alpha\approx 1/137$ is the
fine structure constant. Quantum parameter $\chi_e=(e\hbar/m_e^3c^4)|F_{\mu\nu}p^{\nu}|$ presents the field experienced by the electron in its rest frame normalized to the Schwinger critical field $E_{\rm cr}=1.3\times 10^{16}$ V/cm or $B_{\rm cr}=4.4\times 10^{13}$ G. The laser-plasma interaction enters the QED-dominated region if $\chi_e\gtrsim 1$.

\begin{figure}[t]
\centering
\includegraphics[width=6.3in]{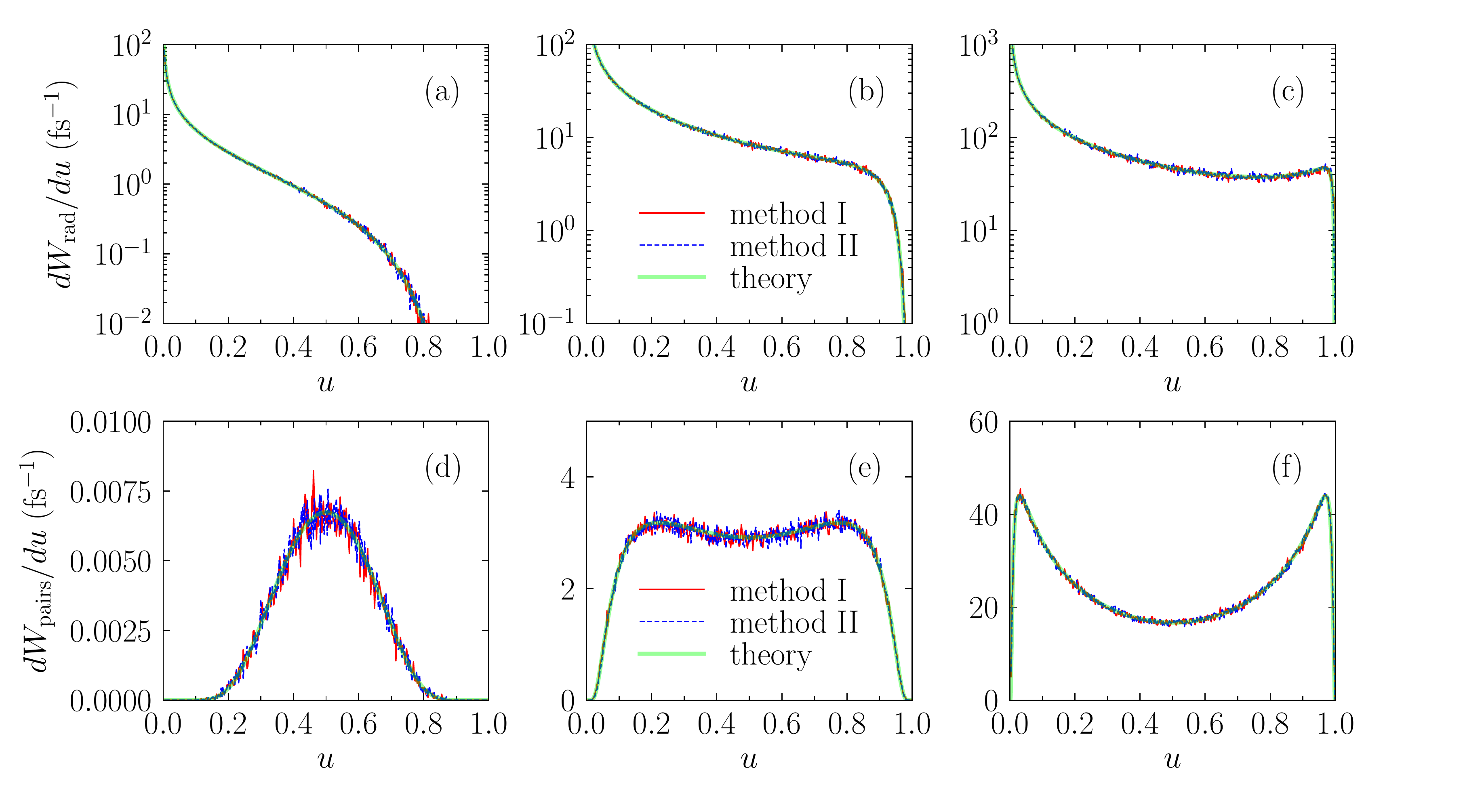}
\caption{\label{fig8} (Upper row) The spectra of emitted photons with $\varepsilon_e=510$ MeV and (a) $\chi_e=0.49$, (b) $\chi_e=4.9$, and (c) $\chi_e=49$, respectively. (Lower row) The spectra of generated positrons with $\varepsilon_\gamma=510$ MeV and (d) $\chi_\gamma=0.49$, (e) $\chi_\gamma=4.9$, and (f) $\chi_\gamma=49$, respectively.}
\end{figure}

To calculate Eq.~\eqref{eqq23}, two different Monte-Carlo methods have been implemented in {\scshape yunic}. Both methods require two uniformly distributed random numbers $r_1$ and $r_2$ to simulate the stochastic photon emission, where $r_1, r_2\in(0,1)$. Method I \cite{Elkina2011prab}: First, a random number $r_1$ is generated to compare with the total radiation probability $W_{\rm rad}$; if $r_1>W_{\rm rad}$, a photon is emitted, and its energy ratio $u_0$ is determined by the other random number $r_2$ according to $\int_{u_{\rm min}}^{u_0}dW_{\rm rad}/du=r_2W_{\rm rad}$ with a low-energy cutoff $u_{\rm min}$. Method II \cite{Elkina2011prab,Gonoskov2015pre}: a photon with an energy ratio of $u_0=r_1^3$ is emitted if $3r_1^2dW_{\rm rad}(u_0)/du>r_2$. The photon spectrum calculated by two methods are both in good agreement with the theoretical spectrum of Eq.~\eqref{eqq23}, as shown in Figs.~\ref{fig8}(a)-(c).

%\subsection{Electron-positron pair production}

%\begin{figure}[t]
%\centering
%\includegraphics[width=6.3in]{fig_positron.pdf}
%\caption{\label{fig8} The spectra of generated positrons with $\gamma_0=1000$ and (a) $\chi_\gamma=0.49$, (b) $\chi_\gamma=4.9$, and (c) $\chi_\gamma=49$, respectively.}
%\end{figure}

Similarly, the spin- and polarized-averaged differential probability of the pair production is written as \cite{Baier1998qed}
\begin{eqnarray}\label{eq2}
\frac{d^2W_{\rm pairs}}{d\varepsilon_{+} dt}&=&\frac{\alpha m^2c^4}{\sqrt{3}\pi \hbar \varepsilon_\gamma^2}\left[\frac{\varepsilon_{+}^2+\varepsilon_{-}^2}{\varepsilon_{+}\varepsilon_{-}}K_{2/3}(y)+\int_{y}^{\infty}K_{1/3}(x)dx \right],
\end{eqnarray}
where $y=2\varepsilon_\gamma^2/(3\chi_\gamma\varepsilon_{+}\varepsilon_{-})$, and $\varepsilon_\gamma$, $\varepsilon_-$ and $\varepsilon_+$ are the energies of the parent $\gamma$ photon, newly created electron and positron, respectively. Another parameter $\chi_\gamma=(e\hbar^2/m_e^3c^4)|F_{\mu\nu}k^{\nu}|$ characterizes the pair production. Two Monte-Carlo methods for the photon emission discussed above are also employed to calculate the pair production in the similar way, which are in good agreement with the theory, as shown in Figs.~\ref{fig8}(d)-(f).

\begin{figure}[t]
\centering
\includegraphics[width=6.3in]{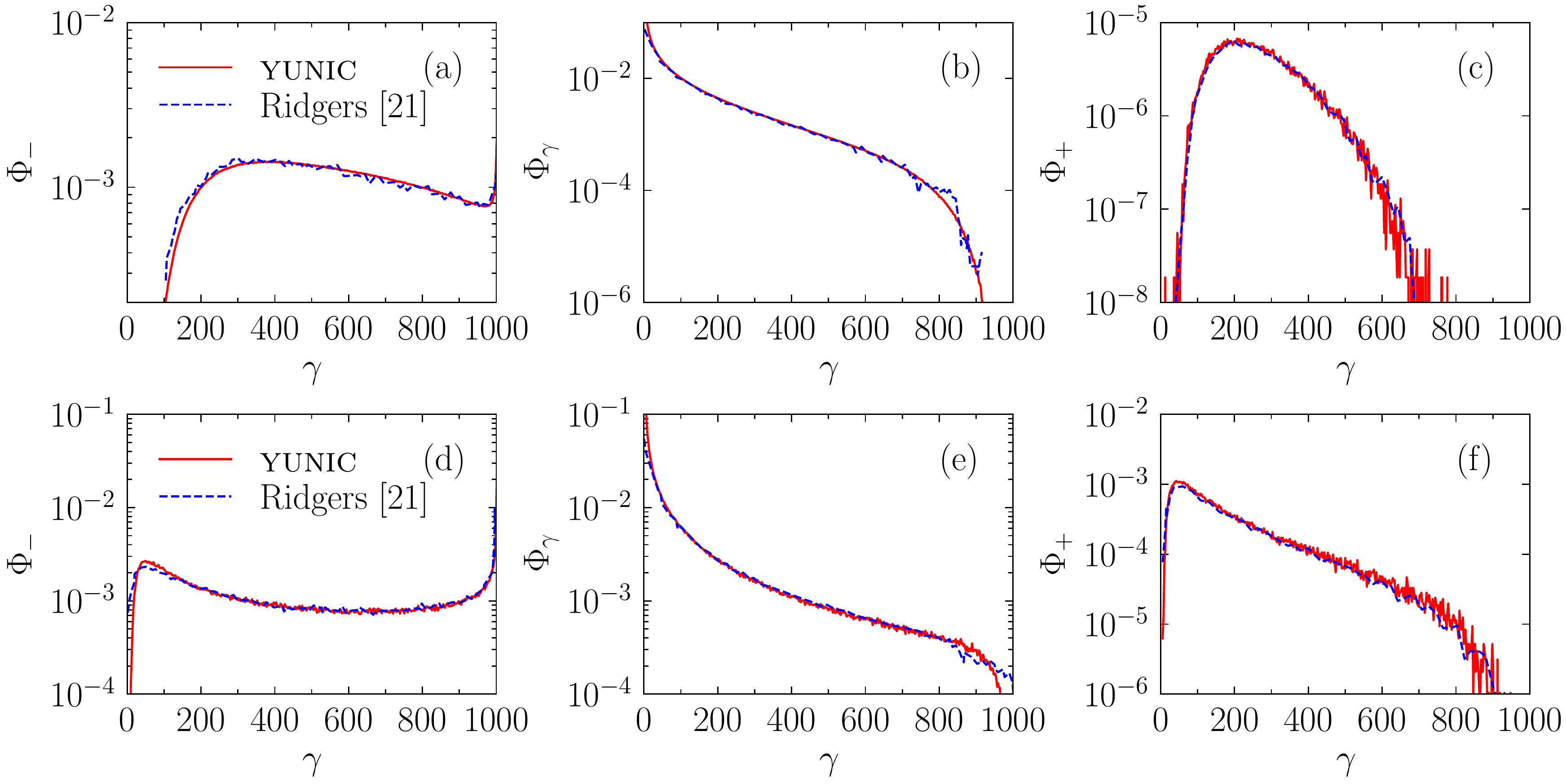}
\caption{\label{fig9} Energy spectra of (a)(d) electrons $\Phi_-$, (b)(e) photons $\Phi_\gamma$, and (c)(f) positrons $\Phi_+$. (a)-(c) and (d)-(f) correspond to strengths of the external magnetic field $B_0=0.001B_{\rm cr}$ and $B_0=0.009B_{\rm cr}$, respectively.}
\end{figure}

Now, we benchmark our QED module in the process of electron-positron cascades with just Method II, where the photon emission, quantum radiation reaction, and pair production are self-consistently included. In Fig.~\ref{fig9}, we consider the same simulation setups as Fig.~2 and Fig.~4 of \cite{Ridgers2014jcp}, where an electron bunch of an initial Lorentz factor $\gamma_0=1000$ are moving under a perpendicularly external magnetic field of a strength of $B_0=0.001B_{\rm cr}$ or $B_0=0.009B_{\rm cr}$. The energy spectra of electrons, photons and positrons both agree well with the results of \cite{Ridgers2014jcp}.

\begin{figure}[t]
\centering
\includegraphics[width=3.3in]{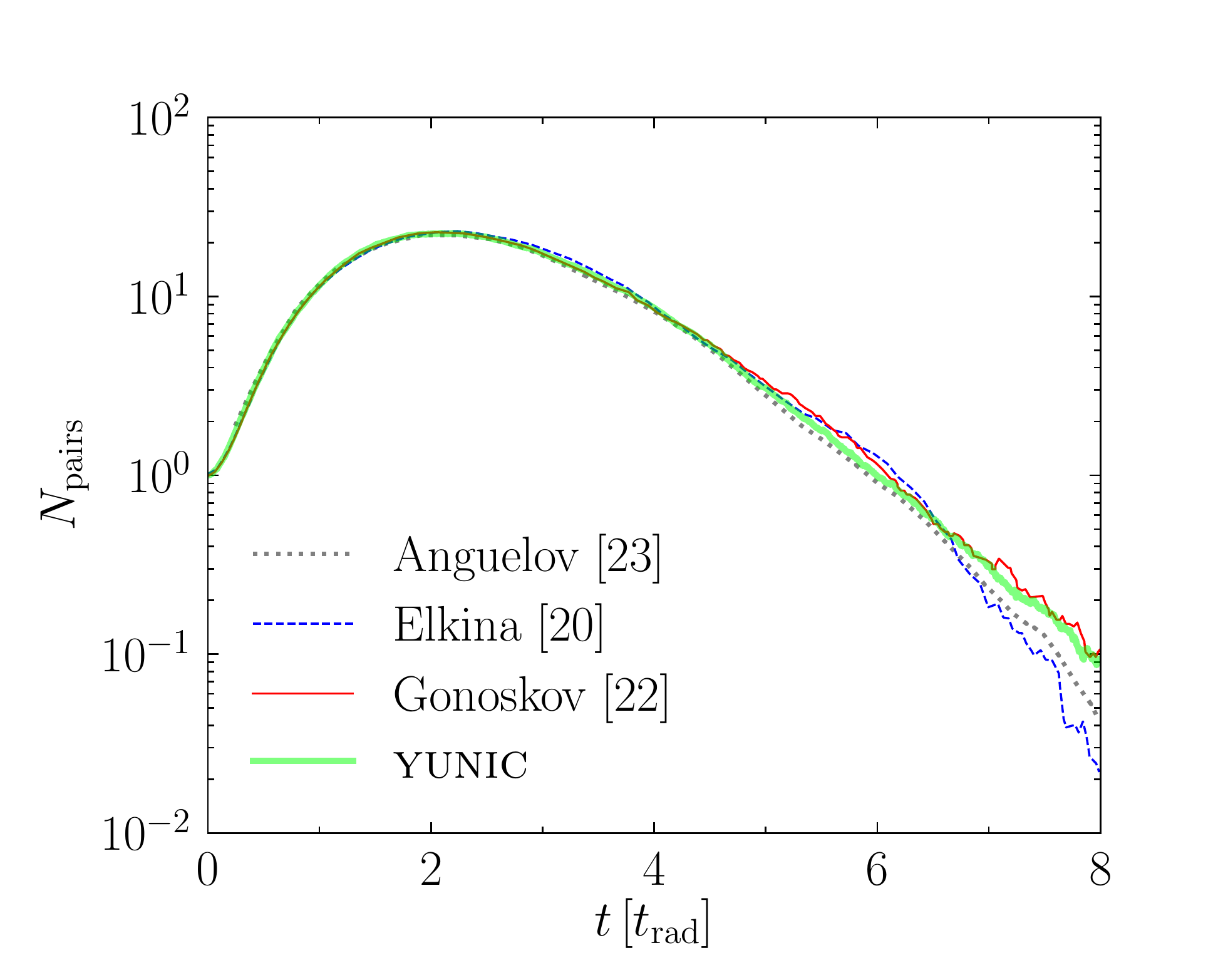}
\caption{\label{fig10} The number of pairs of energies above 200 MeV, where $t_{\rm rad}=1.16\times10^{-16}$ s.}
\end{figure}

In Fig.~\ref{fig10}, we present a typical cascade case to further test our QED module. An electron with an initial Lorentz factor $\gamma_0=2\times 10^5$ is moving under a perpendicularly external magnetic field of a strength of $B_0=0.2B_{\rm cr}$. The total number of electrons and positrons of energies above 100 MeV are countered. The simulation result is shown in Fig.~\ref{fig10}, which is averaged over 2000 simulation runs with different random seeds. Our simulation result is in good agreement with those in \cite{Anguelov1999jpg,Elkina2011prab,Gonoskov2015pre}.

The electron/positron spin and $\gamma$-photon polarization are also implemented into the QED module of {\scshape yunic} \cite{Song2019pra,song2021spin,song2021generation} based on spin- and polarization resolved photon emission and pair production probabilities \cite{Baier1998qed,Li2019prl,Li2020prl1,Li2020prl2}, which have been benchmarked in \cite{song2021spin}. Hence, {\scshape yunic} can be used to investigate spin and polarization related effects in laser-plasma interactions, which are ignored in the previously employed QED-PIC.

\section{Conclusion}
\label{conclusion}
In summary, we have introduced a multi-dimensional PIC code, named {\scshape yunic}. Its core algorithm is described and several benchmarks are preformed. {\scshape Yunic} was employed to investigate the low-frequency whistler waves excited by relativistic laser pulses \cite{Song2020pre} and the spin and polarization effects on the nonlinear Breit-Wheeler pair production in laser-plasma interactions \cite{song2021generation}. In the future, {\scshape yunic} will continue to be used to explore interesting phenomena or physical processes of application prospects and new physical modules may be added for specific problems.

\begin{acknowledgments}
This work was supported by the National Key R\&D Program of China (Grant No.
2018YFA0404801), National Natural Science Foundation of China (Grant Nos. 11775302 and 11721091), the Strategic Priority Research Program of Chinese Academy of Sciences (Grant Nos. XDA25050300, XDA25010300).
\end{acknowledgments}

%\bibliography{yunpic}

\end{document}